\documentclass[12pt]{article}

\usepackage{amsmath}
\usepackage{amssymb}
\usepackage{amsthm}
\usepackage{upref}

\newcommand\reals{\mathbb R}

\title{\bf Faster Evaluation of Multidimensional Integrals}
\author{ A. Papageorgiou \\
J.F. Traub}
\date{Department of Computer Science \\
Columbia University \\
New York, NY 10027 \\ 
June 1997}

\begin{document}

\setcounter{page}{1}
\maketitle

\begin{abstract}
In a recent paper Keister proposed two quadrature rules as alternatives
to Monte Carlo for certain multidimensional integrals and reported his test
results. In earlier work we had shown that the quasi-Monte Carlo method
with generalized Faure points is very effective for a variety of high
dimensional integrals occuring in mathematical finance. In this paper we
report test results of this method on Keister's examples of dimension
$9$ and $25$, and also for examples of dimension $60$, $80$ and $100$.

For the $25$ dimensional integral we achieved accuracy of $10^{-2}$ with
less than $500$ points while the two methods tested by Keister used
more than $220,000$ points. In all of our tests, for $n$ sample points 
we obtained an empirical
convergence rate proportional to $n^{-1}$ rather than the $n^{-1/2}$ of
Monte Carlo.

\end{abstract}

\vskip 2pc
\section{Introduction}
\vskip 1pc

Keister \cite{1} points out that multi-dimensional integrals arise
frequently in many branches of physics. He rules out product rules of
one-dimensional methods because the number of integrand evaluations required
grows exponentially in the number of dimensions. He observes that although 
Monte Carlo (MC) methods are desirable in high dimension, a large number, $n$,
of integrand evaluations can be required since the expected error decreases
as $n^{-1/2}$.

This motivates Keister to seek non-product rules for a certain class of 
integrands defined below. He proposes two quadrature rules, one by
Mc Namee and Stenger (MS) \cite{2}, and a second due to Genz and Patterson
(GP) \cite{3},\cite{4}, which he tests on a specific example of his class of
integrands.

In this paper we report test results on Keister's example using 
quasi-Monte Carlo (QMC) methods. QMC methods evaluate the integrand at
deterministic points in contrast to MC methods which evaluate the integrand
at random points. The deterministic points belong to {\it low discrepancy}
sequences which, roughly speaking, are uniformly spread as we will see
in the next section. Niederreiter \cite{5} is an authoritative monograph
on low discrepancy sequences, their properties, and their applications to
multi-dimensional integration.

The Koksma-Hlawka inequality (see the next section for a precise statement)
states that low discrepancy sequences yield a worst case error for 
multivariate integration bounded by a multiple of
$(\log n)^d/n$, where $n$ is the number of evaluations and $d$ is the
dimension of the integrand. A similar bound on the average error is implied
by Wo\'zniakowski's theorem \cite{6}. The proof of this theorem is
based on concepts and results from information-based complexity \cite{7}.

For $d$ fixed and $n$ large, the error $(\log n)^d/n$ beats the MC error
$n^{-1/2}$. But for $n$ fixed and $d$ large, the $(\log n)^d/n$ factor
looks ominous. Therefore, it was believed that QMC methods should not be 
used for high-dimensional problems; $d=12$ was considered high 
[8, p. 204]. Traub and a then Ph.D. student, Paskov, decided to test the 
efficacy of QMC methods for the valuation of financial derivatives. 
Software construction and testing of QMC methods for financial applications
was began in Fall 1992. The first tests were run on a very difficult
financial derivative in $360$ dimensions, which required $10^5$ floating point
operations per evaluation. Surprisingly, QMC methods consistently beat MC
methods.

The first published announcement was in January 1994 \cite{9}. Details 
appeared in \cite{10}, \cite{11}, \cite{12}. Tests by other researchers
\cite{13}, \cite{14} lead to similar conclusions for the high-dimensional
problems of mathematical finance.

These results are empirical. A number of hypotheses have been advanced to 
explain the observed results. One of these is that, 
due to the discounted value
of money, the financial problems are highly non-isotropic with some
dimensions far more important than others. Perhaps the QMC methods take 
advantage of this. A generally accepted explanation is not
yet available.

Since Keister's test integral is isotropic it provides an example which
is very different than the examples from mathematical finance. To our
surprise the QMC method beat both MC and two other methods tested by
Keister by very convincing margins.

The problems in \cite{1} require the 
computation of a weighted multi-dimensional integral 
\begin{equation}
\int_{\reals^d} f(x) \rho(x)\,dx,
\label {eq:1}
\end{equation}
where $d$ is the dimension
of the problem, $f:\reals^d\to\reals$ is a {\it smooth} function, 
and the weight $\rho(x)$, $x\in\reals^d$ satisfies
\begin{equation}
\rho(x)=\prod_{j=1}^d\eta(x_j),
\label {eq:2}
\end{equation}
with $\eta(-x_j)=\eta(x_j)$, $x_j\in\reals$. Thus, the weight is symmetric
with respect to permutations and changes of sign of the variables.  
The example in \cite{1} (see also \cite{Cap}) is 
\begin{equation}
\int_{\reals^d} \cos(\|x\|) e^{-||x||^2}\, dx,
\label {eq:3}
\end{equation}
where $\|\cdot\|$ denotes the Euclidean norm in $\reals^d$.

The integral in (3) can be reduced, via a change of variable, to a 
one-dimensional integral which can be analytically integrated. As we
will see, the QMC method takes advantage of the dependence on the
norm {\it automatically} and provides a numerical solution with error
similar to a one-dimensional integral.

The QMC method that we test in this paper uses points from the generalized
Faure sequence, which was constructed by Tezuka \cite{15}. We will refer
to it as QMC-GF. This sequence has been very successful in solving problems
of mathematical finance \cite{12}, \cite{14}.

The performance of QMC-GF on the integral (3) is most impressive. For
example, for the $25$-dimensional integral it achieves error $10^{-2}$ 
using less
than $500$ points, far superior to all the other methods. Its error
over the range we tested, which was up to  $10^6$ points, was $c\cdot n^{-1}$,
with $c<110$, $d=9,25,80,60,100$. 
That may be compared with the MC method whose error was 
proportional to $n^{-1/2}$.

We summarize the remainder of this paper. In the next section we provide
a brief introduction to low discrepancy sequences. Test results are given in 
the third section. A summary of our results and future research concludes
the paper.

\vskip 2pc
\section{Low Discrepancy Sequences}
\vskip 1pc

Discrepancy is a measure of deviation from uniformity of a sequence 
of real numbers.
In particular, the discrepancy of $n$ points $x_1,\dots,x_n\in [0,1]^d$, 
$d\geq 1$, is defined by
\begin{equation}
D^{(d)}_n = \sup_{E} \left|\frac{A(E;n)}n - \lambda(E)\right|,
\label {eq:4}
\end{equation}
where the supremum is taken over all the subsets of $[0,1]^d$ of the 
form  $E=[0,t_1)\times\cdots\times [0,t_d)$, 
$0\leq t_j \leq 1$, $1\leq j\leq d$, $\lambda$
denotes the Lebesgue measure, and $A(E;n)$ denotes the number of the $x_j$
that are contained in $E$.
A detailed analysis of low discrepancy sequences can be found in 
\cite{5} and in the references therein.

A sequence $x_1, x_2,\dots$ of points in $[0,1]^d$ is a low discrepancy 
sequence iff 
\begin{equation}
D_n^{(d)}\leq c(d) \frac{(\log n)^d}{n},\; \forall n>1,
\label {eq:5}
\end{equation}
where the constant $c(d)$ depends only on the dimension $d$.
Neiderreiter, see \cite{5}, gives a general method for constructing
$(t,d)$-sequences, 
$t\geq 0$, which are low discrepancy sequences.
The discrepancy of the first $n$ points in
a $(t,d)$-sequence is given by
$$D^{(d)}_n \leq c(t,d,b) \frac{(\log n)^d}{n} + 
O\left( \frac{(\log n)^{d-1}}{n} \right),$$
where $b\geq 2$ is an integer parameter, upon which the sequence depends,
and $c(t,d,b)\approx b^t/d!\cdot (b/2 \log b)^d$.
Hence, the value $t=0$ is desirable.

The generalized Faure sequence \cite{15}
is a $(0,d)$ sequence and is obtained as follows.
For a prime number $b\geq d$ and $n=0,1,\dots$, consider the base $b$ 
representation of $n$, i.e.,
$$n=\sum_{i=0}^{\infty}a_i(n) b^i,$$
where $a_i(n)\in [0,b)$ are integers, $i=0,1,\dots$.
The $j$-th coordinate of the point $x_n$ is then given by
$$x_n^{(j)}=\sum_{k=0}^{\infty} x_{nk}^{(j)} b^{-k-1},\; 1\leq j\leq d,$$
where 
$$x_{nk}^{(j)}=\sum_{s=0}^{\infty} c_{ks}^{(j)} a_s(n).$$
The matrix $C^{(j)}=(c_{ks}^{(j)})$ is called the generator matrix of 
the sequence and is given by $C^{(j)}=A^{(j)} P^{j-1}$, where 
$A^{(j)}$ is a nonsingular lower triangular matrix and $P^{j-1}$ denotes the
$j-1$ power of the Pascal matrix, $1\leq j\leq d$.

We conclude this section by stating the Koksma-Hlawka inequality which
establishes the relationship between low discrepancy sequences and 
multivariate integration, see \cite{5}.
If $f$ is a real function, defined on $[0,1]^d$, of bounded variation, 
$V(f)$, 
in the sense of Hardy and Krause, then for any sequence $x_1,\dots,
x_n\in [0,1)^d$ we have
$$\left| \int_{[0,1]^d} f(x)\, dx - \frac 1n \sum_{i=1}^n f(x_i)\right|
\leq V(f) D_n^{(d)}.$$

\vskip 2pc
\section{Methods and Test Results}
\vskip 1pc

We transform the integral (3) to one over the cube
$[0,1]^d$. We have
\begin{eqnarray}
I_d(\cos)=\int_{\reals^d}\cos(\|x\|)  e^{-\|x\|^2} \,dx & = &
2^{-d/2}\int_{\reals^d}\cos(\|y\|/\sqrt 2)  e^{-\|y\|^2/2} \, dy \\
& = & \pi^{d/2}\int_{\reals^d} \cos(\|y\|/\sqrt 2)  
\frac{e^{-\|y\|^2/2}}{(2\pi)^{d/2}} \, dy \nonumber\\
& = & \pi^{d/2}\int_{[0,1]^d} \cos\left(
\sqrt{\sum_{j=1}^d (\phi^{-1})^2(t_j)/2}\right) \, dt, \nonumber
\label {eq:6}
\end{eqnarray}
where $\phi$ is the cummulative normal distribution function with mean 
$0$ and 
variance $1$,
$$\phi(u)=\frac 1{\sqrt{2\pi}} \int_{-\infty}^u e^{-s^2/2}\, ds,\; 
u\in [-\infty, \infty ].$$

We obtain the $n$ deterministic sample points $x_i=(x_{i1,\dots,x_id})\in 
\reals^d$, $i=1,\dots,n$, by setting $x_{ij}=\phi^{-1}(t_{ij})$, where 
$t_i=(t_{i1},\dots,t_{id})\in [0,1]^d$, $i=1,\dots,n$, are $n$ consecutive
terms of a low discrepancy sequence. Our method, 
$I_{d,n}$, is defined by:
\begin{equation}
I_{d,n}(\cos)=\frac{\pi^{d/2}}n \sum_{i=1}^n \cos\left(
\sqrt{\sum_{j=1}^d (\phi^{-1})^2(t_{i,j})/2}\right).
\label {eq:7}
\end{equation}

Our test problem could be reduced to a one-dimensional integral. We did not do
this because we wanted test how QMC methods
perform on $d$-dimensional integrals. 
As we will see, the empirical rate of convergence of QMC-GF is
$n^{-1}$ which suggests that this method takes advantage of the 
dependence on the norm
automatically without a dimension reducing transformation. 
A method corresponding to (7) can be derived for the
more general integration problem (1) 
with weight function satisfying (2).

We report test results. We used the generalized Faure\footnote{
The generalized Faure and the Sobol' low discrepancy sequences are
included in FINDER, a Columbia University software system, and are available
to researchers upon request by writing the authors.} low
discrepancy sequence \cite{15} to derive the sample points for
the QMC method. We remind the reader that we call this the QMC-GF method.
We compared this method
to the McNamee-Stenger (MS) and Genz-Patterson (GP)
\cite{2}, \cite{3}, \cite{4} methods. We also tested using a 
Monte Carlo method of the form (7), i.e., using randomly generated points
$t_{i,j}$. Hence, we use the same change of variable for QMC-GF and
MC.

The value, $I_d(\cos)$, of the integral (6) is, see \cite{1}, 
$I_9(\cos)=-71.633234291$ and $I_{25}(\cos)=-1.356914\cdot 10^6$. 
We used Mathematica to compute $I_{60}(\cos)=4.89052986\cdot 
10^{14}$, $I_{80}(\cos)=6.78878724\cdot 10^{19}$ and 
$I_{100}(\cos)=4.57024396\cdot 10^{24}$.
We measure the accuracy of an approximation by computing its
relative error (fractional deviation).
We observe the least number of sample points required by
an method to achieve and {\it maintain} a relative error below a 
specified level, e.g. $10^{-3}$, until the end of the simulation. 
We introduced this more conservative
way of assessing the performance of a method in \cite{12}. Thus, we study
the error of an method throughout a simulation. We believe 
that this has advantages over performance reports that are based only on
values at the end of a simulation.

We summarize our findings and then provide some details.
\begin{itemize}
\item The QMC-GF method outperforms the MS and GP
methods for $d=25$.
\item The MS and GP methods are sensitive to the dimension. They
perform quite well for $d=9$ and very poorly for $d=25$.
For example, for $d=25$ and for accuracy of the order $10^{-2}$ these 
methods use some $220,000$ points while the QMC-GF method uses less than
$500$ points.
Therefore, they should only be used when the dimension is relatively low.
\item The QMC-GF method performs well for $d=9$, $25$, $60$, $80$
and $100$.
\item The relative error of the QMC-GF method is bounded by
\begin{equation}
c_d\cdot n^{-1},\; c_d < 110,\;n\leq 10^6,\; d=9, 25, 60, 80, 100.
\label {eq:8}
\end{equation}
Note that this is an empirical conclusion. We write $c_d$ to suggest that,
in principle,
this constant depends on $d$ although we did not see a strong dependence
in our tests.
\item The QMC-GF method achieves relative error  $10^{-2}$
using about $500$ points.
\item The relative error of the MC method is bounded by 
$\beta\cdot n^{-1/2}$
as predicted by the theory. 
\end{itemize}

First we consider the case $d=9$.
The performance of the QMC-GF, and the 
GP methods is comparable 
for accuracy less than $10^{-4}$. 
The relative error of the MS method 
fluctuates about the value $10^{-4}$ for sample
sizes between $36,967$ and $96,745$ points, see [1, Table I],
and is slower than the QMC-GF method since it requires at
least four times as many function evaluations.
(For this level of accuracy the MC method requires 
more than $10^6$ points). 

For $d=25$ the results are striking. The MS and 
GP methods require about $220,000$ points for accuracy
of order $10^{-2}$ while the QMC-GF method requires less
than $500$ points. Table I is from [1, Table II] and exhibits 
the performance of the MS and GP methods. 

\begin{center}
\begin{tabular}{|c|r|c|} \hline
\multicolumn{1}{|c|}{\bf Method} &
 \multicolumn{1}{|c|}{\bf Number of Points} &
 \multicolumn{1}{|c|}{\bf Relative Error} \\ \hline\hline
GP and MS& $1,251$   & $2.00$ \\ \hline
GP       & $19,751$  & $0.40$ \\ \hline
MS       & $20,901$  & $0.75$ \\ \hline
GP       & $227,001$ & $0.06$ \\ \hline
MS       & $244,101$ & $0.07$ \\ \hline 
\multicolumn{3}{c}{\bf Table I. Comparison of MS and GP methods, d=25} 
\end{tabular}
\end{center}

\noindent Table II summarizes the performance of the QMC-GF method.

\begin{center}
\begin{tabular}{|c|r|c|} \hline
\multicolumn{1}{|c|}{\bf Method} &
 \multicolumn{1}{|c|}{\bf Number of Points} &
 \multicolumn{1}{|c|}{\bf Relative Error} \\ \hline\hline
QMC-GF   & $500$       & $10^{-2}$         \\ \hline
QMC-GF   & $1,200$     & $10^{-3}$         \\ \hline   
QMC-GF   & $14,500$    & $5\cdot 10^{-4}$  \\ \hline
QMC-GF   & $214,000$   & $5\cdot 10^{-5}$  \\ \hline 
\multicolumn{3}{c}{\bf Table II. The Quasi-Monte Carlo method, d=25} 
\end{tabular}
\end{center}

\noindent As we mentioned above, we are using a very 
conservative criterion when 
we report relative error.
It takes about 
$219,000$, $490,000$, and many more than $10^6$ points for the MC
method to reach accuracies of
$10^{-3}$, $5\cdot 10^{-4}$, and $5\cdot 10^{-5}$, respectively.

Figure 1 exhibits the relative error of the QMC-GF
method for $d=25$.
The horizontal axis shows the sample size $n$, while
the vertical axis shows the relative error. The horizontal lines depict the
accuracy. 

Figure 2 shows the convergence rate of the QMC-GF method.
We plot the logarithm of the relative error as a function of the logarithm 
of the sample size for $d=25$ and obtain the linear convergence 
summarized in (8).

Recently, Keister \cite{kei97} obtained good results using
a public domain version of the Sobol' low discrepancy sequence.

Keister \cite{1} did not perform tests for $d>25$.
We tested the QMC-GF method for $d=60$, $80$ and $100$ and we found that its
performance is comparable to that of
the lower values of $d$. We did not
find evidence suggesting that its performance suffers as the dimension
grows. This is shown in the empirical error equation (8) and is further
demonstrated in Figure 3, which shows the convergence of QMC-GF for $d=100$.
In particular, in Figure 3 we plot the logarithm of the relative error
as a function of the logarithm of the sample size.

\vskip 2pc
\section{Summary and Future Research}
\vskip 1pc

We have shown that the QMC-GF method beats MC methods and the MS and
GP methods by a wide margin for Keister's $25$-dimensional example. We
have also shown that its good performance is maintained when the dimension 
takes much higher values.
Other high dimensional problems motivated by applications to physics should
be tested.

Extensive testing on a variety of high-dimensional integrals which occur in 
mathematical finance also find QMC methods consistently beating the MC method.
Preliminary results from our tests on high-dimensional integrals arising
from several very different applications again point to the superiority of 
QMC over MC.

The results are empirical. There is currently no theory which explains
why, for a variety of applications, QMC methods are much better than one 
would expect from the Koksma-Hlawka inequality or from Wo\'zniakowski's
theorem.
Finding the theoretical justification for the superiority of QMC methods
for certain classes of integrands is a most important direction of future 
research. 

\vskip 2pc
\section*{Acknowledgments}
\vskip 1pc

We thank Bradley Keister for his comments on a draft of this paper.
We are grateful to Richard Palmer
for directing us to Bradley Keister's paper, and to Henryk Wo\'zniakowski
for his comments on the manuscript.

\clearpage
\setlength{\unitlength}{0.240900pt}
\ifx\plotpoint\undefined\newsavebox{\plotpoint}\fi
\sbox{\plotpoint}{\rule[-0.200pt]{0.400pt}{0.400pt}}%
\begin{picture}(1500,900)(0,0)
\font\gnuplot=cmr10 at 10pt
\gnuplot
\sbox{\plotpoint}{\rule[-0.200pt]{0.400pt}{0.400pt}}%
\put(180.0,82.0){\rule[-0.200pt]{4.818pt}{0.400pt}}
\put(160,82){\makebox(0,0)[r]{0}}
\put(1419.0,82.0){\rule[-0.200pt]{4.818pt}{0.400pt}}
\put(180.0,238.0){\rule[-0.200pt]{4.818pt}{0.400pt}}
\put(160,238){\makebox(0,0)[r]{0.0005}}
\put(1419.0,238.0){\rule[-0.200pt]{4.818pt}{0.400pt}}
\put(180.0,393.0){\rule[-0.200pt]{4.818pt}{0.400pt}}
\put(160,393){\makebox(0,0)[r]{0.001}}
\put(1419.0,393.0){\rule[-0.200pt]{4.818pt}{0.400pt}}
\put(180.0,549.0){\rule[-0.200pt]{4.818pt}{0.400pt}}
\put(160,549){\makebox(0,0)[r]{0.0015}}
\put(1419.0,549.0){\rule[-0.200pt]{4.818pt}{0.400pt}}
\put(180.0,704.0){\rule[-0.200pt]{4.818pt}{0.400pt}}
\put(160,704){\makebox(0,0)[r]{0.002}}
\put(1419.0,704.0){\rule[-0.200pt]{4.818pt}{0.400pt}}
\put(180.0,860.0){\rule[-0.200pt]{4.818pt}{0.400pt}}
\put(160,860){\makebox(0,0)[r]{0.0025}}
\put(1419.0,860.0){\rule[-0.200pt]{4.818pt}{0.400pt}}
\put(304.0,82.0){\rule[-0.200pt]{0.400pt}{4.818pt}}
\put(304,41){\makebox(0,0){5000}}
\put(304.0,840.0){\rule[-0.200pt]{0.400pt}{4.818pt}}
\put(430.0,82.0){\rule[-0.200pt]{0.400pt}{4.818pt}}
\put(430,41){\makebox(0,0){10000}}
\put(430.0,840.0){\rule[-0.200pt]{0.400pt}{4.818pt}}
\put(556.0,82.0){\rule[-0.200pt]{0.400pt}{4.818pt}}
\put(556,41){\makebox(0,0){15000}}
\put(556.0,840.0){\rule[-0.200pt]{0.400pt}{4.818pt}}
\put(682.0,82.0){\rule[-0.200pt]{0.400pt}{4.818pt}}
\put(682,41){\makebox(0,0){20000}}
\put(682.0,840.0){\rule[-0.200pt]{0.400pt}{4.818pt}}
\put(808.0,82.0){\rule[-0.200pt]{0.400pt}{4.818pt}}
\put(808,41){\makebox(0,0){25000}}
\put(808.0,840.0){\rule[-0.200pt]{0.400pt}{4.818pt}}
\put(934.0,82.0){\rule[-0.200pt]{0.400pt}{4.818pt}}
\put(934,41){\makebox(0,0){30000}}
\put(934.0,840.0){\rule[-0.200pt]{0.400pt}{4.818pt}}
\put(1061.0,82.0){\rule[-0.200pt]{0.400pt}{4.818pt}}
\put(1061,41){\makebox(0,0){35000}}
\put(1061.0,840.0){\rule[-0.200pt]{0.400pt}{4.818pt}}
\put(1187.0,82.0){\rule[-0.200pt]{0.400pt}{4.818pt}}
\put(1187,41){\makebox(0,0){40000}}
\put(1187.0,840.0){\rule[-0.200pt]{0.400pt}{4.818pt}}
\put(1313.0,82.0){\rule[-0.200pt]{0.400pt}{4.818pt}}
\put(1313,41){\makebox(0,0){45000}}
\put(1313.0,840.0){\rule[-0.200pt]{0.400pt}{4.818pt}}
\put(1439.0,82.0){\rule[-0.200pt]{0.400pt}{4.818pt}}
\put(1439,41){\makebox(0,0){50000}}
\put(1439.0,840.0){\rule[-0.200pt]{0.400pt}{4.818pt}}
\put(180.0,82.0){\rule[-0.200pt]{303.293pt}{0.400pt}}
\put(1439.0,82.0){\rule[-0.200pt]{0.400pt}{187.420pt}}
\put(180.0,860.0){\rule[-0.200pt]{303.293pt}{0.400pt}}
\put(809,-62){\makebox(0,0){Figure 1. QMC-GF, relative error as a function of the sample size, d=25}}
\put(180.0,82.0){\rule[-0.200pt]{0.400pt}{187.420pt}}
\put(180,393){\usebox{\plotpoint}}
\put(180.0,393.0){\rule[-0.200pt]{303.293pt}{0.400pt}}
\put(180,238){\usebox{\plotpoint}}
\put(180.0,238.0){\rule[-0.200pt]{303.293pt}{0.400pt}}
\multiput(190.58,738.46)(0.493,-12.748){23}{\rule{0.119pt}{10.008pt}}
\multiput(189.17,759.23)(13.000,-301.229){2}{\rule{0.400pt}{5.004pt}}
\multiput(203.58,434.34)(0.492,-7.217){21}{\rule{0.119pt}{5.700pt}}
\multiput(202.17,446.17)(12.000,-156.169){2}{\rule{0.400pt}{2.850pt}}
\multiput(215.58,285.75)(0.493,-1.171){23}{\rule{0.119pt}{1.023pt}}
\multiput(214.17,287.88)(13.000,-27.877){2}{\rule{0.400pt}{0.512pt}}
\multiput(228.58,260.00)(0.493,2.875){23}{\rule{0.119pt}{2.346pt}}
\multiput(227.17,260.00)(13.000,68.130){2}{\rule{0.400pt}{1.173pt}}
\multiput(241.58,302.84)(0.492,-9.242){21}{\rule{0.119pt}{7.267pt}}
\multiput(240.17,317.92)(12.000,-199.918){2}{\rule{0.400pt}{3.633pt}}
\multiput(253.58,113.37)(0.493,-1.290){23}{\rule{0.119pt}{1.115pt}}
\multiput(252.17,115.68)(13.000,-30.685){2}{\rule{0.400pt}{0.558pt}}
\multiput(266.58,85.00)(0.492,5.106){21}{\rule{0.119pt}{4.067pt}}
\multiput(265.17,85.00)(12.000,110.559){2}{\rule{0.400pt}{2.033pt}}
\multiput(278.58,204.00)(0.493,4.065){23}{\rule{0.119pt}{3.269pt}}
\multiput(277.17,204.00)(13.000,96.215){2}{\rule{0.400pt}{1.635pt}}
\multiput(291.58,282.32)(0.493,-7.514){23}{\rule{0.119pt}{5.946pt}}
\multiput(290.17,294.66)(13.000,-177.658){2}{\rule{0.400pt}{2.973pt}}
\multiput(304.00,117.60)(1.651,0.468){5}{\rule{1.300pt}{0.113pt}}
\multiput(304.00,116.17)(9.302,4.000){2}{\rule{0.650pt}{0.400pt}}
\multiput(316.58,121.00)(0.493,7.237){23}{\rule{0.119pt}{5.731pt}}
\multiput(315.17,121.00)(13.000,171.105){2}{\rule{0.400pt}{2.865pt}}
\multiput(329.00,302.93)(1.378,-0.477){7}{\rule{1.140pt}{0.115pt}}
\multiput(329.00,303.17)(10.634,-5.000){2}{\rule{0.570pt}{0.400pt}}
\multiput(342.58,299.00)(0.492,3.641){21}{\rule{0.119pt}{2.933pt}}
\multiput(341.17,299.00)(12.000,78.912){2}{\rule{0.400pt}{1.467pt}}
\multiput(354.58,346.03)(0.493,-11.638){23}{\rule{0.119pt}{9.146pt}}
\multiput(353.17,365.02)(13.000,-275.017){2}{\rule{0.400pt}{4.573pt}}
\multiput(367.58,90.00)(0.492,3.167){21}{\rule{0.119pt}{2.567pt}}
\multiput(366.17,90.00)(12.000,68.673){2}{\rule{0.400pt}{1.283pt}}
\multiput(379.00,164.59)(1.378,0.477){7}{\rule{1.140pt}{0.115pt}}
\multiput(379.00,163.17)(10.634,5.000){2}{\rule{0.570pt}{0.400pt}}
\multiput(392.58,159.64)(0.493,-2.757){23}{\rule{0.119pt}{2.254pt}}
\multiput(391.17,164.32)(13.000,-65.322){2}{\rule{0.400pt}{1.127pt}}
\multiput(405.58,99.00)(0.492,5.709){21}{\rule{0.119pt}{4.533pt}}
\multiput(404.17,99.00)(12.000,123.591){2}{\rule{0.400pt}{2.267pt}}
\multiput(417.58,227.50)(0.493,-1.250){23}{\rule{0.119pt}{1.085pt}}
\multiput(416.17,229.75)(13.000,-29.749){2}{\rule{0.400pt}{0.542pt}}
\multiput(430.00,200.58)(0.496,0.492){21}{\rule{0.500pt}{0.119pt}}
\multiput(430.00,199.17)(10.962,12.000){2}{\rule{0.250pt}{0.400pt}}
\multiput(442.00,210.92)(0.539,-0.492){21}{\rule{0.533pt}{0.119pt}}
\multiput(442.00,211.17)(11.893,-12.000){2}{\rule{0.267pt}{0.400pt}}
\multiput(455.58,200.00)(0.493,5.334){23}{\rule{0.119pt}{4.254pt}}
\multiput(454.17,200.00)(13.000,126.171){2}{\rule{0.400pt}{2.127pt}}
\multiput(468.58,335.00)(0.492,1.487){21}{\rule{0.119pt}{1.267pt}}
\multiput(467.17,335.00)(12.000,32.371){2}{\rule{0.400pt}{0.633pt}}
\multiput(480.58,342.00)(0.493,-8.545){23}{\rule{0.119pt}{6.746pt}}
\multiput(479.17,356.00)(13.000,-201.998){2}{\rule{0.400pt}{3.373pt}}
\multiput(493.58,151.51)(0.492,-0.625){21}{\rule{0.119pt}{0.600pt}}
\multiput(492.17,152.75)(12.000,-13.755){2}{\rule{0.400pt}{0.300pt}}
\multiput(505.58,139.00)(0.493,2.638){23}{\rule{0.119pt}{2.162pt}}
\multiput(504.17,139.00)(13.000,62.514){2}{\rule{0.400pt}{1.081pt}}
\multiput(518.58,206.00)(0.493,0.774){23}{\rule{0.119pt}{0.715pt}}
\multiput(517.17,206.00)(13.000,18.515){2}{\rule{0.400pt}{0.358pt}}
\multiput(531.58,226.00)(0.492,1.099){21}{\rule{0.119pt}{0.967pt}}
\multiput(530.17,226.00)(12.000,23.994){2}{\rule{0.400pt}{0.483pt}}
\multiput(543.58,237.92)(0.493,-4.224){23}{\rule{0.119pt}{3.392pt}}
\multiput(542.17,244.96)(13.000,-99.959){2}{\rule{0.400pt}{1.696pt}}
\multiput(556.58,141.77)(0.493,-0.853){23}{\rule{0.119pt}{0.777pt}}
\multiput(555.17,143.39)(13.000,-20.387){2}{\rule{0.400pt}{0.388pt}}
\multiput(569.58,123.00)(0.492,0.798){21}{\rule{0.119pt}{0.733pt}}
\multiput(568.17,123.00)(12.000,17.478){2}{\rule{0.400pt}{0.367pt}}
\multiput(581.00,142.59)(1.378,0.477){7}{\rule{1.140pt}{0.115pt}}
\multiput(581.00,141.17)(10.634,5.000){2}{\rule{0.570pt}{0.400pt}}
\multiput(594.58,138.84)(0.492,-2.392){21}{\rule{0.119pt}{1.967pt}}
\multiput(593.17,142.92)(12.000,-51.918){2}{\rule{0.400pt}{0.983pt}}
\multiput(606.58,91.00)(0.493,0.536){23}{\rule{0.119pt}{0.531pt}}
\multiput(605.17,91.00)(13.000,12.898){2}{\rule{0.400pt}{0.265pt}}
\multiput(619.58,101.77)(0.493,-0.853){23}{\rule{0.119pt}{0.777pt}}
\multiput(618.17,103.39)(13.000,-20.387){2}{\rule{0.400pt}{0.388pt}}
\multiput(632.58,83.00)(0.492,1.358){21}{\rule{0.119pt}{1.167pt}}
\multiput(631.17,83.00)(12.000,29.579){2}{\rule{0.400pt}{0.583pt}}
\multiput(644.58,115.00)(0.493,1.329){23}{\rule{0.119pt}{1.146pt}}
\multiput(643.17,115.00)(13.000,31.621){2}{\rule{0.400pt}{0.573pt}}
\multiput(657.00,147.92)(0.496,-0.492){21}{\rule{0.500pt}{0.119pt}}
\multiput(657.00,148.17)(10.962,-12.000){2}{\rule{0.250pt}{0.400pt}}
\multiput(669.58,129.94)(0.493,-2.043){23}{\rule{0.119pt}{1.700pt}}
\multiput(668.17,133.47)(13.000,-48.472){2}{\rule{0.400pt}{0.850pt}}
\multiput(682.58,85.00)(0.493,1.964){23}{\rule{0.119pt}{1.638pt}}
\multiput(681.17,85.00)(13.000,46.599){2}{\rule{0.400pt}{0.819pt}}
\multiput(695.58,135.00)(0.492,0.582){21}{\rule{0.119pt}{0.567pt}}
\multiput(694.17,135.00)(12.000,12.824){2}{\rule{0.400pt}{0.283pt}}
\multiput(707.58,149.00)(0.493,1.210){23}{\rule{0.119pt}{1.054pt}}
\multiput(706.17,149.00)(13.000,28.813){2}{\rule{0.400pt}{0.527pt}}
\multiput(720.58,171.79)(0.493,-2.400){23}{\rule{0.119pt}{1.977pt}}
\multiput(719.17,175.90)(13.000,-56.897){2}{\rule{0.400pt}{0.988pt}}
\multiput(733.58,119.00)(0.492,2.349){21}{\rule{0.119pt}{1.933pt}}
\multiput(732.17,119.00)(12.000,50.987){2}{\rule{0.400pt}{0.967pt}}
\multiput(745.58,169.88)(0.493,-1.131){23}{\rule{0.119pt}{0.992pt}}
\multiput(744.17,171.94)(13.000,-26.940){2}{\rule{0.400pt}{0.496pt}}
\multiput(758.58,141.96)(0.492,-0.798){21}{\rule{0.119pt}{0.733pt}}
\multiput(757.17,143.48)(12.000,-17.478){2}{\rule{0.400pt}{0.367pt}}
\multiput(770.58,122.39)(0.493,-0.972){23}{\rule{0.119pt}{0.869pt}}
\multiput(769.17,124.20)(13.000,-23.196){2}{\rule{0.400pt}{0.435pt}}
\multiput(783.00,101.58)(0.652,0.491){17}{\rule{0.620pt}{0.118pt}}
\multiput(783.00,100.17)(11.713,10.000){2}{\rule{0.310pt}{0.400pt}}
\multiput(796.58,107.96)(0.492,-0.798){21}{\rule{0.119pt}{0.733pt}}
\multiput(795.17,109.48)(12.000,-17.478){2}{\rule{0.400pt}{0.367pt}}
\multiput(808.00,90.93)(0.950,-0.485){11}{\rule{0.843pt}{0.117pt}}
\multiput(808.00,91.17)(11.251,-7.000){2}{\rule{0.421pt}{0.400pt}}
\multiput(821.58,85.00)(0.492,1.099){21}{\rule{0.119pt}{0.967pt}}
\multiput(820.17,85.00)(12.000,23.994){2}{\rule{0.400pt}{0.483pt}}
\multiput(833.58,111.00)(0.493,0.774){23}{\rule{0.119pt}{0.715pt}}
\multiput(832.17,111.00)(13.000,18.515){2}{\rule{0.400pt}{0.358pt}}
\put(846,130.67){\rule{3.132pt}{0.400pt}}
\multiput(846.00,130.17)(6.500,1.000){2}{\rule{1.566pt}{0.400pt}}
\multiput(859.58,124.80)(0.492,-2.090){21}{\rule{0.119pt}{1.733pt}}
\multiput(858.17,128.40)(12.000,-45.402){2}{\rule{0.400pt}{0.867pt}}
\multiput(871.58,83.00)(0.493,1.091){23}{\rule{0.119pt}{0.962pt}}
\multiput(870.17,83.00)(13.000,26.004){2}{\rule{0.400pt}{0.481pt}}
\multiput(884.00,111.60)(1.797,0.468){5}{\rule{1.400pt}{0.113pt}}
\multiput(884.00,110.17)(10.094,4.000){2}{\rule{0.700pt}{0.400pt}}
\multiput(897.58,115.00)(0.492,0.841){21}{\rule{0.119pt}{0.767pt}}
\multiput(896.17,115.00)(12.000,18.409){2}{\rule{0.400pt}{0.383pt}}
\multiput(909.00,133.92)(0.652,-0.491){17}{\rule{0.620pt}{0.118pt}}
\multiput(909.00,134.17)(11.713,-10.000){2}{\rule{0.310pt}{0.400pt}}
\multiput(922.58,125.00)(0.492,1.056){21}{\rule{0.119pt}{0.933pt}}
\multiput(921.17,125.00)(12.000,23.063){2}{\rule{0.400pt}{0.467pt}}
\multiput(934.58,144.86)(0.493,-1.448){23}{\rule{0.119pt}{1.238pt}}
\multiput(933.17,147.43)(13.000,-34.430){2}{\rule{0.400pt}{0.619pt}}
\multiput(947.58,113.00)(0.493,1.250){23}{\rule{0.119pt}{1.085pt}}
\multiput(946.17,113.00)(13.000,29.749){2}{\rule{0.400pt}{0.542pt}}
\multiput(960.58,137.94)(0.492,-2.047){21}{\rule{0.119pt}{1.700pt}}
\multiput(959.17,141.47)(12.000,-44.472){2}{\rule{0.400pt}{0.850pt}}
\multiput(972.58,97.00)(0.493,2.162){23}{\rule{0.119pt}{1.792pt}}
\multiput(971.17,97.00)(13.000,51.280){2}{\rule{0.400pt}{0.896pt}}
\put(190.0,780.0){\rule[-0.200pt]{0.400pt}{19.272pt}}
\multiput(997.58,149.16)(0.493,-0.734){23}{\rule{0.119pt}{0.685pt}}
\multiput(996.17,150.58)(13.000,-17.579){2}{\rule{0.400pt}{0.342pt}}
\multiput(1010.00,133.59)(1.123,0.482){9}{\rule{0.967pt}{0.116pt}}
\multiput(1010.00,132.17)(10.994,6.000){2}{\rule{0.483pt}{0.400pt}}
\multiput(1023.58,135.96)(0.492,-0.798){21}{\rule{0.119pt}{0.733pt}}
\multiput(1022.17,137.48)(12.000,-17.478){2}{\rule{0.400pt}{0.367pt}}
\multiput(1035.00,120.61)(2.695,0.447){3}{\rule{1.833pt}{0.108pt}}
\multiput(1035.00,119.17)(9.195,3.000){2}{\rule{0.917pt}{0.400pt}}
\multiput(1048.58,123.00)(0.493,1.250){23}{\rule{0.119pt}{1.085pt}}
\multiput(1047.17,123.00)(13.000,29.749){2}{\rule{0.400pt}{0.542pt}}
\multiput(1061.58,152.09)(0.492,-0.755){21}{\rule{0.119pt}{0.700pt}}
\multiput(1060.17,153.55)(12.000,-16.547){2}{\rule{0.400pt}{0.350pt}}
\multiput(1073.00,135.93)(0.950,-0.485){11}{\rule{0.843pt}{0.117pt}}
\multiput(1073.00,136.17)(11.251,-7.000){2}{\rule{0.421pt}{0.400pt}}
\multiput(1086.58,123.91)(0.492,-1.746){21}{\rule{0.119pt}{1.467pt}}
\multiput(1085.17,126.96)(12.000,-37.956){2}{\rule{0.400pt}{0.733pt}}
\multiput(1098.58,89.00)(0.493,1.369){23}{\rule{0.119pt}{1.177pt}}
\multiput(1097.17,89.00)(13.000,32.557){2}{\rule{0.400pt}{0.588pt}}
\multiput(1111.00,122.92)(0.590,-0.492){19}{\rule{0.573pt}{0.118pt}}
\multiput(1111.00,123.17)(11.811,-11.000){2}{\rule{0.286pt}{0.400pt}}
\multiput(1124.58,109.82)(0.492,-0.841){21}{\rule{0.119pt}{0.767pt}}
\multiput(1123.17,111.41)(12.000,-18.409){2}{\rule{0.400pt}{0.383pt}}
\multiput(1136.58,93.00)(0.493,0.536){23}{\rule{0.119pt}{0.531pt}}
\multiput(1135.17,93.00)(13.000,12.898){2}{\rule{0.400pt}{0.265pt}}
\multiput(1149.00,105.92)(0.600,-0.491){17}{\rule{0.580pt}{0.118pt}}
\multiput(1149.00,106.17)(10.796,-10.000){2}{\rule{0.290pt}{0.400pt}}
\multiput(1161.00,95.92)(0.590,-0.492){19}{\rule{0.573pt}{0.118pt}}
\multiput(1161.00,96.17)(11.811,-11.000){2}{\rule{0.286pt}{0.400pt}}
\put(1174,84.17){\rule{2.700pt}{0.400pt}}
\multiput(1174.00,85.17)(7.396,-2.000){2}{\rule{1.350pt}{0.400pt}}
\multiput(1187.58,84.00)(0.492,1.315){21}{\rule{0.119pt}{1.133pt}}
\multiput(1186.17,84.00)(12.000,28.648){2}{\rule{0.400pt}{0.567pt}}
\multiput(1199.00,113.92)(0.539,-0.492){21}{\rule{0.533pt}{0.119pt}}
\multiput(1199.00,114.17)(11.893,-12.000){2}{\rule{0.267pt}{0.400pt}}
\multiput(1212.00,101.93)(0.728,-0.489){15}{\rule{0.678pt}{0.118pt}}
\multiput(1212.00,102.17)(11.593,-9.000){2}{\rule{0.339pt}{0.400pt}}
\multiput(1225.00,92.92)(0.543,-0.492){19}{\rule{0.536pt}{0.118pt}}
\multiput(1225.00,93.17)(10.887,-11.000){2}{\rule{0.268pt}{0.400pt}}
\multiput(1237.58,83.00)(0.493,0.655){23}{\rule{0.119pt}{0.623pt}}
\multiput(1236.17,83.00)(13.000,15.707){2}{\rule{0.400pt}{0.312pt}}
\multiput(1250.00,100.61)(2.472,0.447){3}{\rule{1.700pt}{0.108pt}}
\multiput(1250.00,99.17)(8.472,3.000){2}{\rule{0.850pt}{0.400pt}}
\multiput(1262.58,99.90)(0.493,-0.814){23}{\rule{0.119pt}{0.746pt}}
\multiput(1261.17,101.45)(13.000,-19.451){2}{\rule{0.400pt}{0.373pt}}
\multiput(1275.58,82.00)(0.493,0.695){23}{\rule{0.119pt}{0.654pt}}
\multiput(1274.17,82.00)(13.000,16.643){2}{\rule{0.400pt}{0.327pt}}
\multiput(1288.00,98.93)(0.874,-0.485){11}{\rule{0.786pt}{0.117pt}}
\multiput(1288.00,99.17)(10.369,-7.000){2}{\rule{0.393pt}{0.400pt}}
\multiput(1300.58,93.00)(0.493,0.695){23}{\rule{0.119pt}{0.654pt}}
\multiput(1299.17,93.00)(13.000,16.643){2}{\rule{0.400pt}{0.327pt}}
\multiput(1313.00,109.94)(1.651,-0.468){5}{\rule{1.300pt}{0.113pt}}
\multiput(1313.00,110.17)(9.302,-4.000){2}{\rule{0.650pt}{0.400pt}}
\multiput(1325.00,107.61)(2.695,0.447){3}{\rule{1.833pt}{0.108pt}}
\multiput(1325.00,106.17)(9.195,3.000){2}{\rule{0.917pt}{0.400pt}}
\multiput(1338.58,106.65)(0.493,-0.893){23}{\rule{0.119pt}{0.808pt}}
\multiput(1337.17,108.32)(13.000,-21.324){2}{\rule{0.400pt}{0.404pt}}
\put(1351,86.67){\rule{2.891pt}{0.400pt}}
\multiput(1351.00,86.17)(6.000,1.000){2}{\rule{1.445pt}{0.400pt}}
\put(1363,86.67){\rule{3.132pt}{0.400pt}}
\multiput(1363.00,87.17)(6.500,-1.000){2}{\rule{1.566pt}{0.400pt}}
\put(1376,85.67){\rule{3.132pt}{0.400pt}}
\multiput(1376.00,86.17)(6.500,-1.000){2}{\rule{1.566pt}{0.400pt}}
\multiput(1389.00,86.58)(0.600,0.491){17}{\rule{0.580pt}{0.118pt}}
\multiput(1389.00,85.17)(10.796,10.000){2}{\rule{0.290pt}{0.400pt}}
\multiput(1401.00,94.94)(1.797,-0.468){5}{\rule{1.400pt}{0.113pt}}
\multiput(1401.00,95.17)(10.094,-4.000){2}{\rule{0.700pt}{0.400pt}}
\multiput(1414.00,90.93)(0.669,-0.489){15}{\rule{0.633pt}{0.118pt}}
\multiput(1414.00,91.17)(10.685,-9.000){2}{\rule{0.317pt}{0.400pt}}
\put(985.0,152.0){\rule[-0.200pt]{2.891pt}{0.400pt}}
\put(1426.0,83.0){\rule[-0.200pt]{3.132pt}{0.400pt}}
\end{picture}

\vskip 1.5in

\input fig2.tex

\clearpage
\input fig3.tex

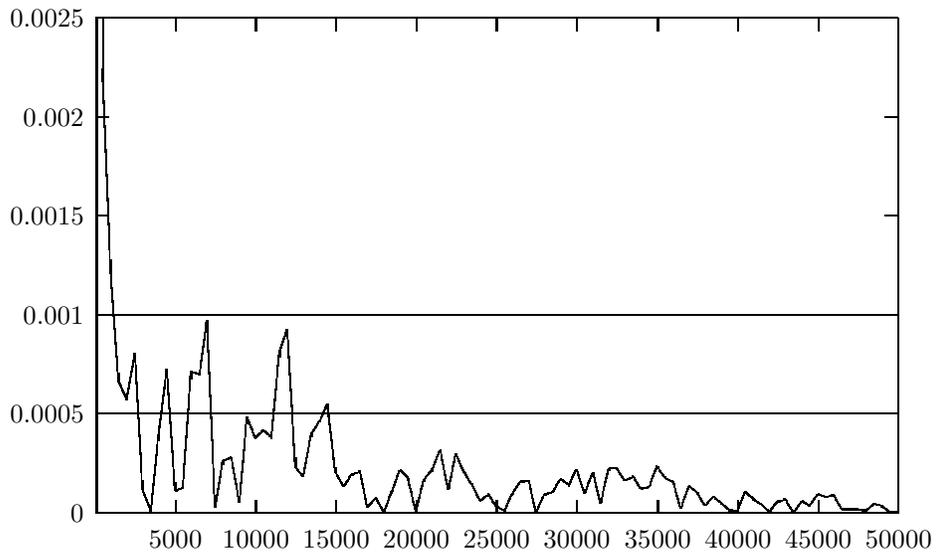
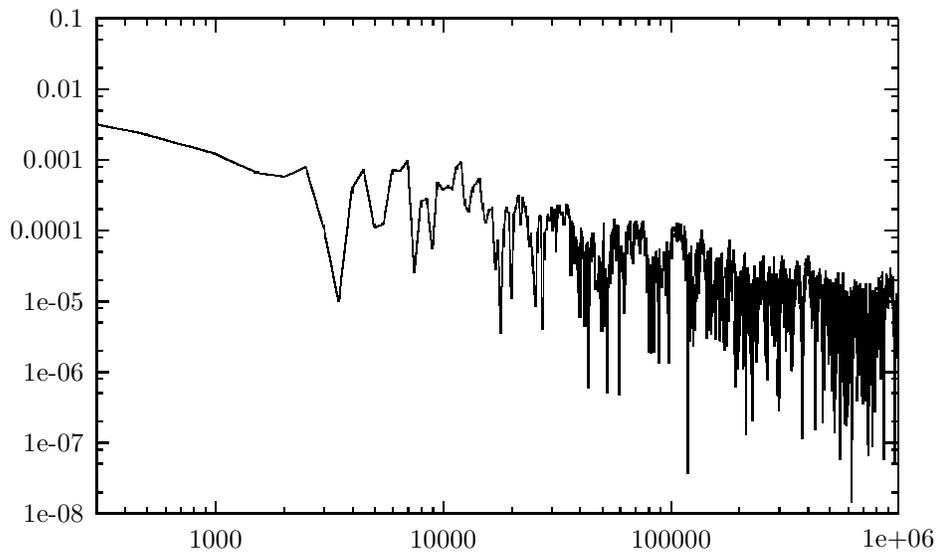
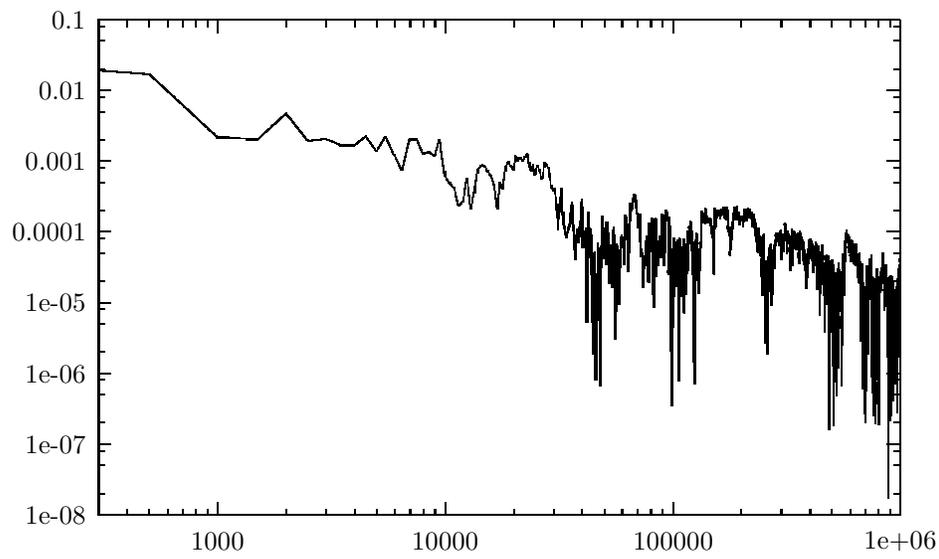
\end{document}